\documentclass[aps,prl,twocolumn,floatfix]{revtex4-2}

\usepackage{amsmath}
\usepackage{amssymb}
\usepackage{graphicx}
\usepackage{bm}
\usepackage{color}
\usepackage{times}
\usepackage[final]{microtype}

\begin{document}
\newcommand{\mal}{\mathcal{L}}
\newcommand{\ra}{\rangle}
\newcommand{\la}{\langle}
\newcommand{\mbf}[1]{\mathbf{#1}}
\newcommand{\mat}[1]{\mathcal{#1}}
\newcommand{\Mob}{M{\"o}bius }

\title{Band-Structure-Independent Topology from Nonsymmorphic Wannier Complexes}

\author{Qinghua He$^{1}$}
\thanks{These two authors contributed equally}

\author{Jie Zhang$^{2,3}$}
\thanks{These two authors contributed equally}

\author{Shengdan Tao$^{2}$}
\author{Hai-yao Deng$^{4}$}
\author{Qifeng Liang$^{2}$}
\email{qfliang@usx.edu.cn}
\author{Wenlong Gao$^{4}$}
\email{wgao@eitech.edu.cn}
\author{Feng Liu$^{2,5,6}$}
\email{ruserzzz@gmail.com}

\affiliation{$^{1}$Institute of High Pressure Physics, School of Physical Science and Technology, Ningbo University, Ningbo 315-211, China}
\affiliation{$^{2}$Department of Physics, Shaoxing University, Shaoxing 312-000, China}
\affiliation{$^{3}$Department of Physics, Qufu Normal University, Qufu 273100, China}
\affiliation{$^{4}$School of Physics and Astronomy, Cardiff University, 5 The Parade, Cardiff CF24 3AA, United Kingdom}
\affiliation{$^{5}$Eastern Institute of Technology, Ningbo, China}
\affiliation{$^{6}$Department of Nanotechnology for Sustainable Energy, School of Science and Technology, Kwansei Gakuin University, Gakuen 2-1, Sanda 669-1337, Japan}

\begin{abstract}
  Nonsymmorphic symmetries can enforce band connectivity that obstructs a single-band Wannier description. We show that a fractional translation $\mal$ connecting distinct high-symmetry Wyckoff positions generically renders the Wannier center of an individual band gauge ill-defined, requiring a symmetry-enforced multiband object---a Wannier complex. We formulate a real-space topological classification of Wannier complexes and show that, when $\mal$ is combined with certain point-group symmetries (notably $C_4$ and $C_3$), all symmetry-allowed Wannier-complex configurations carry a nontrivial quantized total electric polarization. This yields boundary phenomena that persist across symmetry-preserving deformations of the Hamiltonian, including parameter regimes with and without bulk gaps. We demonstrate the mechanism in minimal tight-binding models exhibiting M{\"o}bius-twisted Wilson-loop structures and higher-order corner modes, and propose experimental signatures in a dielectric photonic crystal and a first-principles electronic platform octa-graphene, accompanied by a three-dimensional extension.
\end{abstract}

\maketitle
\section{Introduction}
Topology and symmetry are foundational concepts in modern condensed matter physics~\cite{Wen1990, Xie2013}, whose combination has yielded topological materials possessing robust properties like helical edge states~\cite{Hatsugai1993,Kane2005,Bernevig2006,Po2017, Bradlyn2017}.
Yet in most existing topological materials, the nontrivial response is tied to specific band structures, often described through band inversion or a particular pattern of band ordering, so that continuous tune of parameters such as hopping amplitudes and onsite energies even in a symmetry-preserving way moves the system between topological and trivial regimes.
This fact creates a conceptual tension: topological responses are expected to be robust under symmetry-preserving deformations, but the very mechanisms used to realize them are typically sensitive to microscopic band structure sdetails.
This tension raises a more fundamental question: can crystalline symmetries enforce topological phenomena that persist irrespective of the detailed band structure across insulating, semimetallic, and even metallic regimes, without invoking strong interactions or magnetic fluxes?

A useful way to see why this is nontrivial is through the standard Wannier-based formulation of band topology~\cite{Ryu2010, Chiu2016, Ahn2020}.
In band theory, a topological obstruction can be diagnosed by the impossibility of constructing exponentially localized Wannier functions for an isolated set of bands~\cite{Yu2011}, or by the symmetry-constrained locations of Wannier centers relative to atomic sites~\cite{Khalaf2021,Ma2023, Xu2024}.
These criteria are powerful precisely because they connect momentum-space structure to real-space representation.
However, in common use, they still depend on how energy bands are isolated and ordered.
Crucially, topology is often assigned to a chosen band or band subspace, and changes in the band structure can change which subspace is relevant~\cite{Xu2023PRL}.
A paradigmatic example is the Su-Schrieffer-Heeger (SSH) model and its analogues~\cite{Su1979, Liu2017, Liu2023}, where band inversion, driven by hopping texture tuning, leads to a quantized bulk polarization~\cite{Zak1989, King1993, Liu2023}. Yet, this topology arises only when two distinct band structures are contrasted. When the two bands with complementary Wannier character or band ordering are grouped together, the system becomes trivial and hence the topology there is fragile and relative~\cite{Po2018}.
In this sense, much of conventional band topology is band-structure dependent~\cite{Kruthoff2017}, even though the invariants are well defined and robust within a fixed range.

\begin{figure}[t]
  \leavevmode
  \begin{center}
    \leavevmode
    \includegraphics[clip=true,width=0.99\columnwidth]{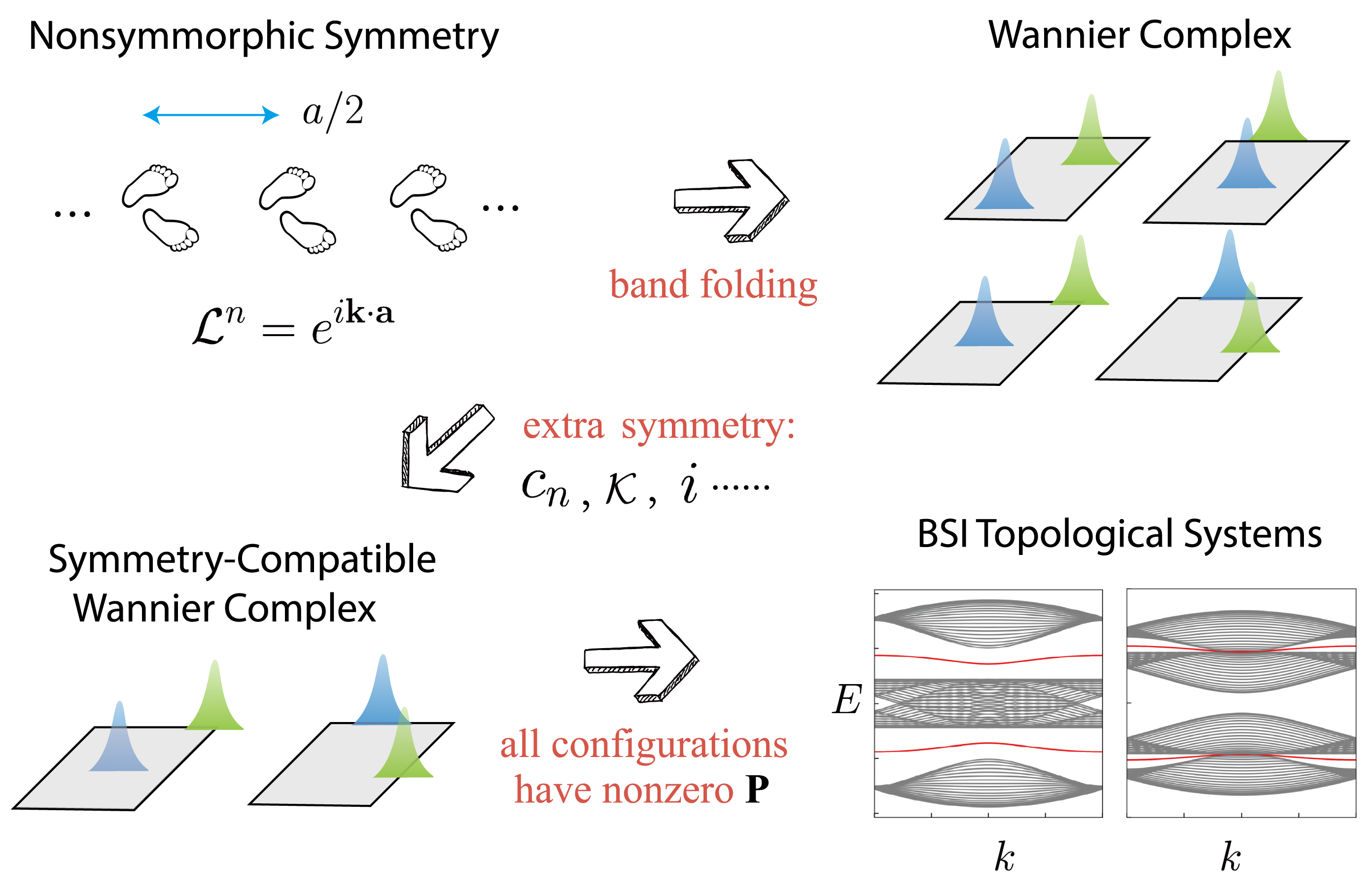}
  \end{center}
  \caption{Schematic of the core concept for BSI topological systems.
    In a system endowed with nonsymmorphic symmetry manifested as fractional translation $\mal$, band folding yields the emergence of a Wannier complex: a group of Wannier-function centers denoted as $(\mbf{p}_1,\mbf{p}_2,\cdots,\mbf{p}_m)$ connected by $\mal$.
    By introducing additional symmetries such as point-group symmetry, the configuration of the Wannier complex is further constrained. 
    In some cases, all symmetry-compatible configurations of Wannier complexes have a nontrivial quantized total polarization, which results in BSI topological systems that are robust across symmetry-preserving deformations.
  }
\end{figure}

Here we present a route to a stronger notion of symmetry-driven topology by exploiting a specific class of nonsymmorphic symmetry that we denote by $\mal$.
Operationally, $\mal$ is a space-group element, $\mal=\{e|\mbf{L}\}$, with $e$ the identity point operation, $\mbf{L}$ a fractional translation connecting distinct high-symmetry Wyckoff positions, and $\mbf{L}^m=T_\mbf{a}$.
This operation exists when the orbitals form a non-primitive basis such that $\mbf{L}$ maps one site in the basis to another equivalent site.
A key consequence of $\mathcal{L}$ is not merely an enforced band degeneracy, but a more basic obstruction in the Wannier description: the Wannier center of a single band becomes gauge ill-defined.
As detailed in a minimal one-dimensional (1D) setting discussed later, applying $\mbf{L}$ shifts the Wannier center of an $\mal$-eigenstate by half a lattice constant, contradicting the expected gauge-invariant monodromy.
Instead of assigning a symmetry-consistent Wannier center to an individual band, one is forced to treat a set of Wannier centers collectively as a single composite object.
We call this object a \emph{Wannier complex}, a symmetry-enforced group of Wannier centers $(\mbf{p}_1,\mbf{p}_2,\cdots,\mbf{p}_m)$ connected by $\mbf{L}$ (and, in general, by additional crystalline symmetries).

Once topology is reformulated in terms of Wannier complexes rather than individual bands, the topological classification problem becomes intrinsically geometric and symmetry-driven.
The central claim we develop is that, for certain combinations of $\mal$ with point-group symmetries, the allowed Wannier-complex configurations are so tightly constrained that all symmetry-compatible configurations carry a nontrivial quantized total electric polarization.
In that case, the existence of nontrivial boundary responses is fixed at the level of symmetry and real-space representation, rather than by a particular band-inversion pattern.
This is the sense in which we use the term band-structure-independent (BSI) topology: the real-space topological character of the Wannier complex is universal across symmetry-preserving deformations of the Hamiltonian. We illustrate the path to BSI topology in Fig.~1.

In this paper, we develop this BSI framework systematically and identify symmetry settings where it becomes universal.
We demonstrate explicitly the BSI topology with $C_4$- and $C_3$-symmetric toy models, where $\mal$ enforces Wannier complexes of order two and three, respectively.
These settings yield robust boundary phenomena including edge modes (with M{\"o}bius-like structure in suitable boundary constructions)~\cite{Shiozaki2015} and higher-order corner states~\cite{Song2017, Benalcazar2019, Liu2019}.
To connect the principles with realistic platforms, we propose an experimentally feasible dielectric photonic crystal realization of the $\mal+C_4$ mechanism, and verify directional interface modes and localized corner modes via full-wave finite-element simulations.
Notably, shifting the point source by the $\mbf{L}$ toggles excitations between counterpropagating channels at fixed frequency.
We also identify a 2D electronic material platform, octa-graphene, exhibiting the corresponding symmetry setting and present ribbon spectra with edge-localized states consistent with the Wannier-complex diagnosis.

The paper is organized as follows. Section II establishes the foundational properties of systems with $\mal$ symmetry using a one-dimensional two-band model and introduces the Wannier-complex viewpoint.
Sections III–V extend the construction to 2D with different point groups, highlighting the symmetry conditions under which Wannier-complex polarization becomes universally nontrivial and detailing the associated edge, Wilson-loop, and higher-order boundary signatures.
We then discuss experimental proposals and material realization and conclude with a 3D extension and outlook.
\section{Wannier complexes in a 1D Case}
We begin with a 1D lattice model equipped with nonsymmorphic symmetry.
One general form of a two-band model is given by
\begin{equation}
  H_k=p_k\sigma_z+q_k\sigma_x+\lambda_k\sigma_y,
\end{equation}
where $p_k$, $q_k$, and $\lambda_k$ are continuous real functions of the momentum $k$, and $\sigma_i$ ($i=x,y,z$) are the Pauli matrices.
The nonsymmorphic operator in the 1D case is given by
\begin{equation}
  \mathcal{L}=
  \begin{pmatrix}
    0&e^{ika} \\
    1&0
  \end{pmatrix},
\end{equation}
where $a$ is the lattice constant, and the basis consists of the two nonequivalent sublattices of the 1D model defined within a unit cell.
Note that $\mathcal{L}^2=e^{ika}$ due to the Bloch theorem. The expression for $\mathcal{L}$ is not unique; different forms are related by gauge transformations.

A schematic of one possible 1D lattice model endowed with $\mal$ symmetry is depicted in Fig.~2(a). The two distinct sublattices are indicated by colors, and the hopping parameter is $\lambda_n=\int dk (q_k-i\lambda_k)e^{-ikna}$ between two sublattices separated by $n$ unit cells from left to right.
Intrinsically, the lattice model has a fractional translation $\mbf{L}=a/2$ as $\mal$.

As $\mal^2=e^{ika}$, the eigenvalue of $\mal$ is $\pm e^{ika/2}$.
The corresponding eigenvectors are
\begin{equation}
  |\pm\rangle=\frac{1}{\sqrt{2}}
  \begin{pmatrix}
    1\\
    \pm e^{-ik/2}
  \end{pmatrix},
\end{equation}
which are also eigenvectors of $H_k$ as $[H_k,\mal]=0$.
Note that the periodicity of $|\pm\rangle$ is $4\pi$ rather than $2\pi$, because $|+,k+2\pi\rangle=|-,k\rangle$, which forms a twist in momentum space analogous to a M{\"o}bius strip.

Due to this twist structure, the Wannier center of a single eigenstate is gauge ill-defined, which can be seen explicitly by applying $\mal$ on $|+\rangle$ as
\begin{equation}
  \begin{split}
    W^\prime_c&=\frac{1}{2\pi}\oint dk \langle +'| i\partial_k |+'\rangle\\
    &=\frac{1}{2\pi}\oint dk\langle+'|i\partial_ke^{ik/2}+e^{ik/2}i\partial_k|+\rangle \\
    &=W_c-\frac{1}{2},
  \end{split}
\end{equation}
where $|+'\rangle=\mal|+\rangle$.
From Eq.~(4) we observe that, by applying $\mal$, the Wannier center of $|+\rangle$ shifts by half a lattice constant, which contradicts the monodromy and gauge invariance of Wannier centers.
This contradiction can naturally extend to higher dimensions as seen in later discussion for 2D and 3D cases.
Because of the gauge ill-defined Wannier center for a single band under $\mal$, the Wannier centers of these two eigenvectors should be considered together, such as $(p_1,p_2)$, characterized by, for example, their summation. We call such Wannier centers, which appear in a pair (or a group) enforced by $\mal$ and other symmetries, a Wannier complex.
Formally, the sum of a Wannier complex is given by the trace of the integral of the non-Abelian Berry connection over the first Brillouin zone (BZ), such as $\mbf{P}=\sum_i^m \int_\text{BZ} \mbf{A}_{i,i}\cdot d\mbf{k}$ with $\mbf{A}_{i,j}=\langle u_i|i\partial_\mbf{k}|u_j\rangle$ and $u_i$ the periodic part of the Bloch function for the $i$th band.

\begin{figure}[t]
  \leavevmode
  \begin{center}
    \leavevmode
    \includegraphics[clip=true,width=0.99\columnwidth]{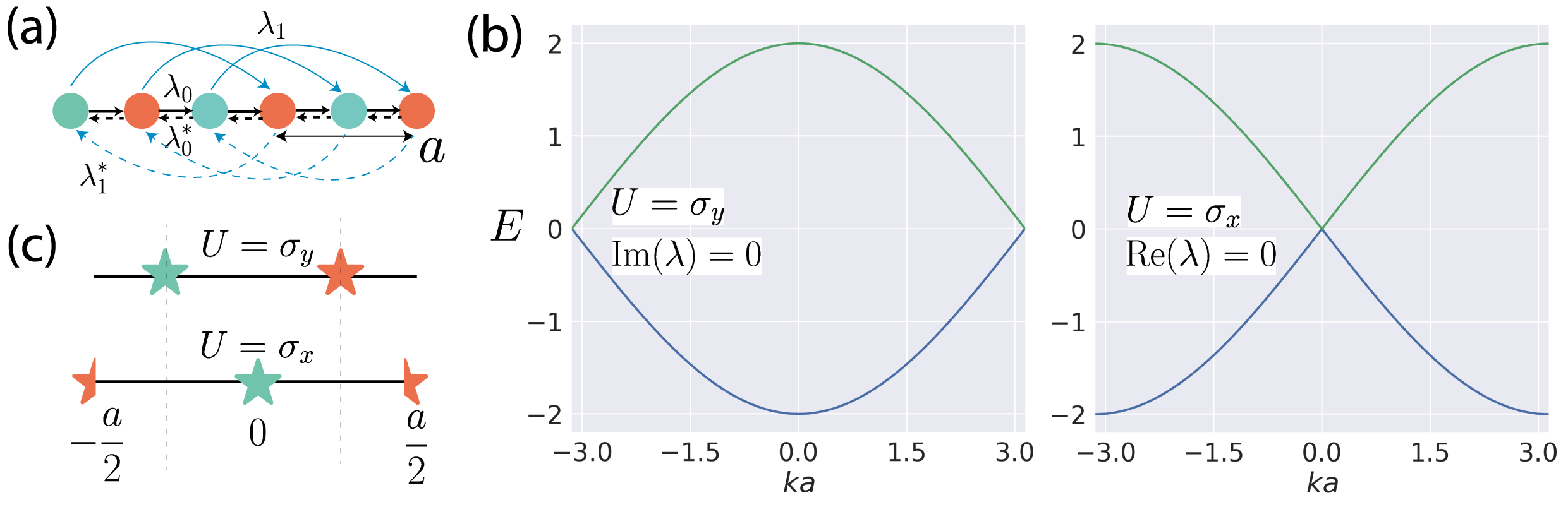}
  \end{center}
  \caption{(a) Lattice model of 1D system with $\mal$ symmetry. The lattice constant is $a$. There are two types of sublattices, and the hopping from left to right is $\lambda_n$, and $\lambda_n^*$ for the opposite direction with $n$ indicating the distance. The hopping pattern leads to the formation of a fractional translation $\mbf{L}$.
    (b) Energy band structures of the 1D lattice for real and imaginary hopping amplitudes $\lambda$.
    (c) Two configurations of the Wannier complex corresponding to the band structures in (b), highlighting the different total electric polarization configurations for each case.
  }
\end{figure}
Now we investigate possible configurations of the Wannier complex of the 1D two-band model under crystalline symmetries, such as inversion symmetry.
First, under the constraint of $\mal$ symmetry, we have $\mal^{-1}H_k\mal=H_k$, producing~\cite{Zhao2016}
\begin{equation}
  \begin{split}
    p_k &=-p_k, \\
    (q_k+i\lambda_k)e^{ika} &= q_k-i\lambda_k.
  \end{split}
\end{equation}
To lowest order, considering only nearest-neighbor hopping, we have $q_k-i\lambda_k=\lambda_0+\lambda_0^*e^{ika}$ with $\lambda_0=\lambda$.
Second, under the constraint of inversion symmetry, implemented by the operator $\mat{P}=U\hat{i}$ with $U$ being a unitary operator satisfying $U^2=1$ and $\hat{i}$ flipping the sign of momentum $k$, we have
\begin{equation}
  \lambda^*=\pm \lambda,
\end{equation}
where the $\pm$ sign corresponds to $U=\sigma_y$ and $U=\sigma_x$.

Solving the eigenproblem of $H_k$ for these two cases, we obtain two configurations of the Wannier complex for real and imaginary $\lambda$s.
Figure 2(b) displays the band structures of the two cases, and Fig.~2(c) displays the two types of corresponding Wannier complex configurations.
For real $\lambda$, it has a total electric polarization of 0 featuring a degenerate point at $k=\pi$, while for imaginary $\lambda$, the total electric polarization is $1/2$ featuring a degeneracy at $k=0$. For a complex $\lambda$ without inversion symmetry, the total electric polarization varies between 0 and $1/2$.

\subsection{Topological Classification of Wannier Complexes}
We can use the Wannier complex to construct a representation for the underlying crystalline symmetry group that includes $n$-fold rotations $C_n$ and lattice translations as well as $\mal$.
The bases of such a representation $\Gamma$ are the Wannier centers (defined in the same unit cell) comprising the Wannier complex, denoted by $(\mbf{p}^\Gamma_1, \mbf{p}^\Gamma_2, \dots, \mbf{p}^\Gamma_m)$.
They should then satisfy
\begin{equation}
  g \mbf{p}^\Gamma_i = \sum_j D^{\Gamma}_{ji}(g)\mbf{p}^\Gamma_j,
\end{equation}
where $g$ is an element of the symmetry group, $D^{\Gamma}(g)$ is the matrix of $g$ in the representation.
In the 1D case with $U=\sigma_y$, $p_1+1/2=p_2$ and $-p_2=p_1$, and we have $(p_1,p_2)=(-a/4,a/4)$.
On the other hand, with $U=\sigma_x$, $-p_{1,2} \text{ mod } a=p_{1,2}$ and $p_1+1/2=p_2$, and we have $(p_1,p_2)=(0,a/2)$.

It is noted that $(\mbf{p}_1,\mbf{p}_2,\cdots,\mbf{p}_m)$ forms a product space and a symmetry operation simply permutes the Wannier centers and hence the bases $\mathbf{p}_i$ ($g$ either fixes $\mbf{p}_i$ or switches it with its partner).
In other words, the order of $\mbf{p}_i$s is not important.
Thus, the topological classification of $(\mbf{p}_1,\mbf{p}_2,\cdots,\mbf{p}_m)$ is given by its quotient space by the permutation group, that is, the symmetric group $\text{Sym}^m(X)$, where $X$ is the base space, such as a circle $S^1$ and a torus $T^2$ in 1D and 2D systems.
For example, in the 1D two bands model, we have $\text{Sym}^2(S^1)\simeq\text{M{\"o}bius strip}$, whose fundamental group is $\pi_1[\text{Sym}^2(S^1)]=Z$, such as the winding number.

In general, the classification of the symmetric group and its associated Wannier complex can be carried out first by its order (rank) $m$. Within the same order, it can be further classified by its degree, which can be characterized by the total electric polarization $\mathbf{P}=\sum^m_i \mbf{p}_i$ (first Stiefel–Whitney index)~\cite{Liu2017,Ahn2019PRB, Xue2023}, the Chern class, and other topological invariants~\cite{Shiozaki2015,Kim2019PRB}. The complete topological classification of Wannier complexes in two dimensions is given by vector bundles over an elliptic curve~\cite{Atiyah1957}.

\section{BSI topology with $C_4$ Point-Group Symmetry}
We extend our discussion from 1D to 2D.
In 2D systems, we have $\mal^n=e^{i\mathbf{k}\cdot(l\mathbf{a}_1+m\mathbf{a}_2)}$, where $\mathbf{a}_i$ are the primitive vectors, and $l,m$ each take the value 1 or 0.
For simplicity, we consider $\mal$ symmetry that connects two special Wyckoff positions, which are left invariant by the identity operation and at least one other operation of the space group.
Besides $\mal$, we focus on point-group symmetry $C_n$ for $n=2,3,4,6$.
In the following discussion, we show that, with $C_4$ point-group symmetry, the total electric polarization of the corresponding Wannier complex is always nonzero for a specific $\mal$, associated with M{\"o}bius edge states across insulating, semimetallic, and metallic phases.

There are four Wyckoff positions for the $C_4$ point group, consisting of three special Wyckoff positions and one general Wyckoff position.
The three special Wyckoff positions are $1a$, $1b$, and $2c$, where the number denotes the multiplicity, as displayed in Fig.~3(a).
There are three choices of $\mal$, namely $l=0, m=1, n=2$, $l=1,m=0, n=2$, and $l=1,m=1,n=2$, corresponding to the fractional translations $(a/2, 0)$, $(0, a/2)$, and $(a/2,a/2)$, respectively.

For $\mbf{L}=(a/2,0)$ and $\mbf{L}=(0,a/2)$, we can easily show that the order of the Wannier complex is four, for which the total electric polarization vanishes.
To find all the configurations of the Wannier complex under symmetries of $\mal$ and $C_n$, one can enumerate all the initial Wannier centers at different Wyckoff positions, including both general and special ones.
For example, for $\mbf{L}=(a/2,0)$, we set the initial Wannier center at $(0,0)$; then by acting the symmetry operation on the initial center, we obtain all the configurations as $\mbf{p}_1=(0,0)$, $\mbf{p}_2=(0,a/2)$, $\mbf{p}_3=(a/2,0)$ and $\mbf{p}_4=(a/2,a/2)$.

\begin{figure}[t]
  \leavevmode
  \begin{center}
    \leavevmode
    \includegraphics[clip=true,width=0.99\columnwidth]{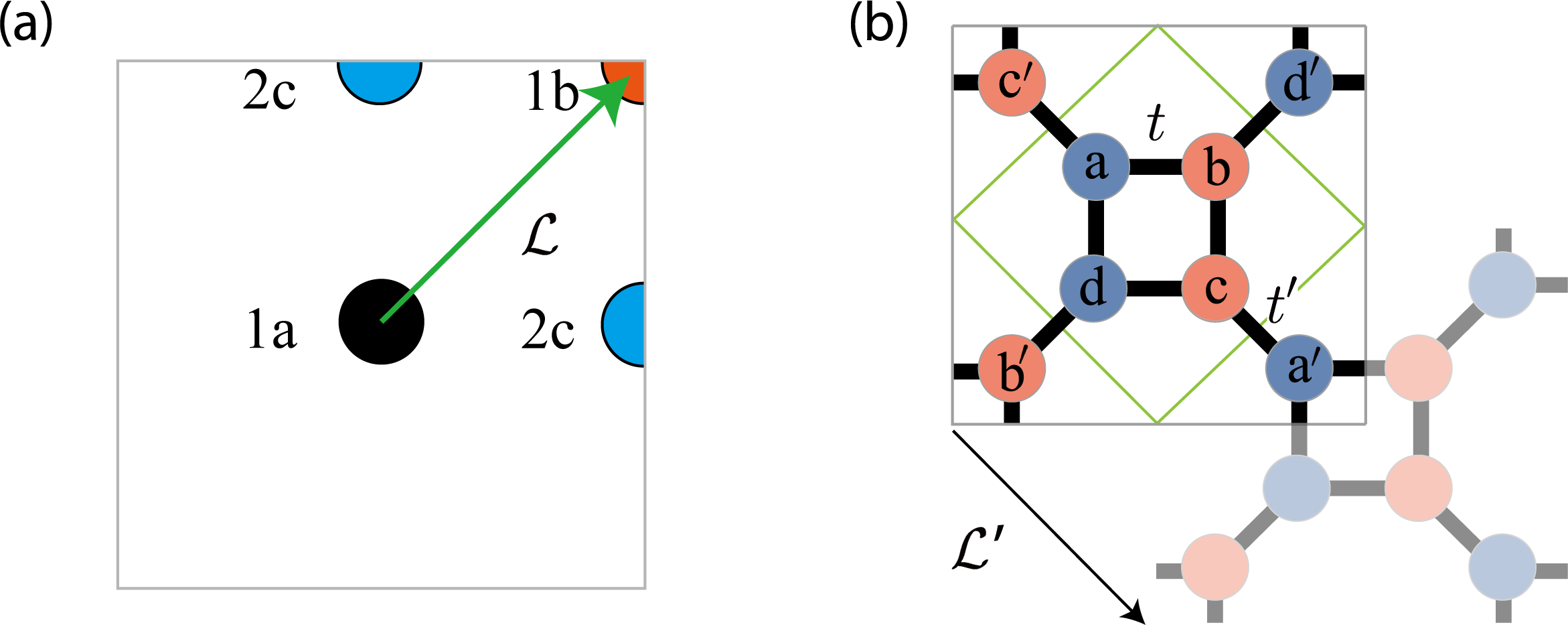}
  \end{center}
  \caption{(a) The Wyckoff positions in a $C_4$-symmetric lattice, with the three special Wyckoff positions marked 1a, 1b, and 2c, as illustrated in the diagram.
    (b) Minimal model of a $C_4$-symmetric BSI topological system.
    Colors serve as an eye guide for equivalent sites connected by $\mbf{L}=(a/2,a/2)$, such as $a$ and $a^\prime$.
  }
\end{figure}
\subsection{BSI Topological System}
For $\mbf{L}=(a/2,a/2)$, we have two possible configurations of the Wannier complex. One has bases $\mbf{p}_1=(0,0)$ and $\mbf{p}_2=(a/2,a/2)$, while the other has bases $\mbf{p}_1=(a/2,0)$ and $\mbf{p}_2=(0,a/2)$.
The order of this complex is two less than the degeneracy of the general Wyckoff position in a $C_4$-symmetric lattice.
Both configurations have nontrivial electric polarization, and edge states are expected across all symmetry-compatible insulating, semi-metallic, and metallic phases, irrespective of a specific band structure.
This yields a BSI topological system.

To validate the discussion about the BSI topological system, we consider a square $C_4$ symmetric lattice model equipped with $\mbf{L}=(a/2,a/2)$, as displayed in Fig.~3(b).
There are four nonequivalent sublattices in the primitive unit cell as marked within the light diamond shape in Fig.~3(b).
The colors serve as an eye guide for those sublattices connected by $\mal$ in the doubled unit cell in Fig.~3(b).
There are two types of hoppings, namely $t$ and $t^\prime$.
It is noted that $\mathcal{L}$ operations along certain directions are equivalent, such as $\mathcal{L}^\prime=(a/2, -a/2)$, due to the lattice period $a$.

The Hamiltonian of the proposed $C_4$ model in terms of the primitive cell can be written as
\begin{equation}
  H_\mbf{k}=
  \begin{pmatrix}
    \mu&t&t^\prime e^{-i \frac{k_x-k_y}{2}} & t\\
    t &-\mu & t & t^\prime e^{i\frac{k_x+k_y}{2}}\\
    t^\prime e^{i\frac{k_x-k_y}{2}} &t &\mu &t \\
    t & t^\prime e^{-i\frac{k_x+k_y}{2}} & t &-\mu
  \end{pmatrix},
\end{equation}
where $t$, $t^\prime$ are the intracell and intercell hopping amplitudes, respectively, $\pm \mu$ are the onsite potentials of the sublattices, which are zero for $C_4$ symmetric systems, and the bases are $|a\rangle$,$|b\rangle$,$|c\rangle$,$|d\rangle$ as denoted in Fig.~3(b).
Considering $\mal$ and $C_4$ symmetries, we can infer that the model of eight sublattices in Fig.~3(b) is a minimal model for the BSI topological system.

\begin{figure}[t]
  \leavevmode
  \begin{center}
    \leavevmode
    \includegraphics[clip=true,width=0.99\columnwidth]{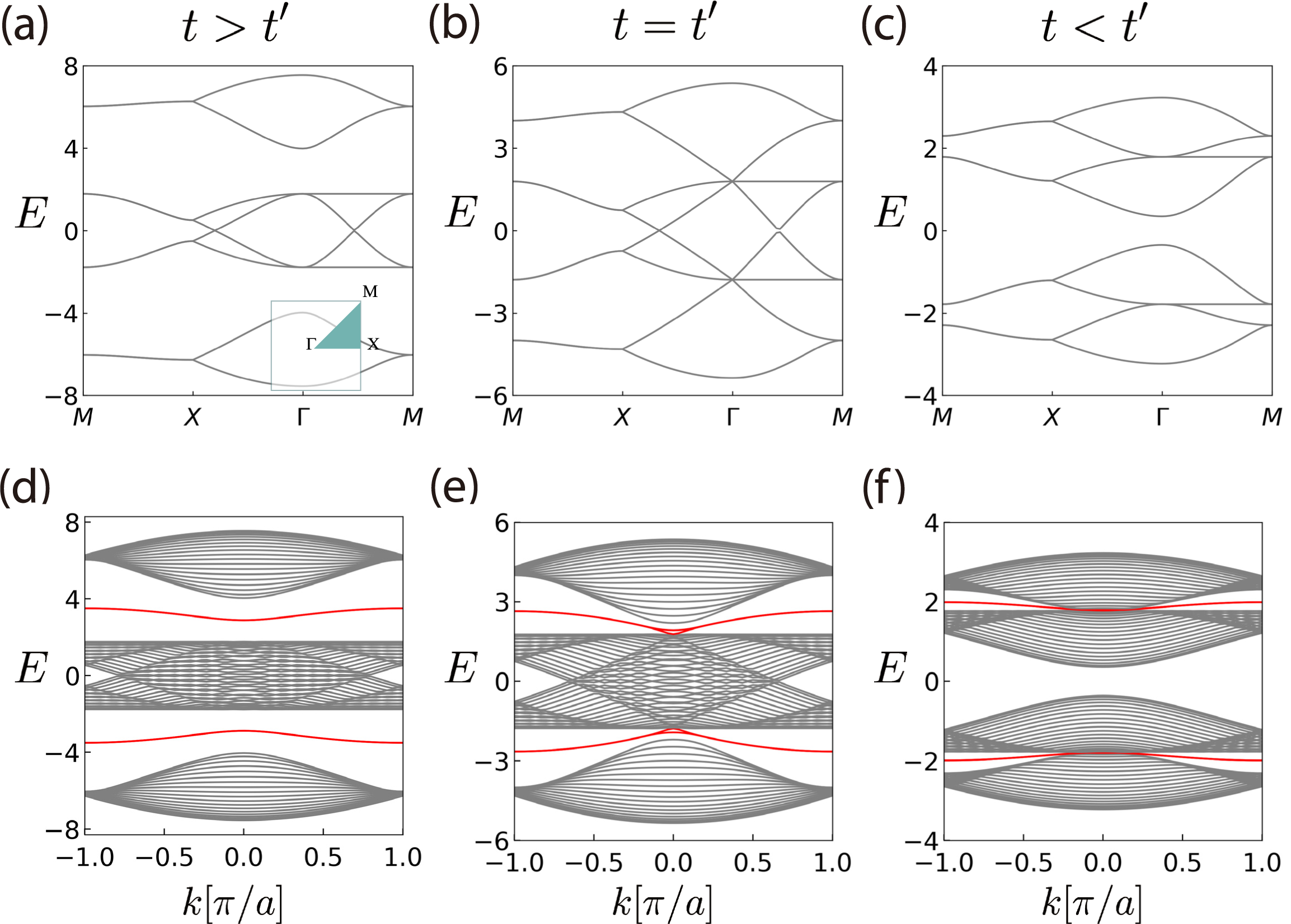}
  \end{center}
  \caption{Bulk energy spectra of the $C_4$-symmetric model with $\mathcal{L}$ symmetry for different phases: (a) insulating phase for $|t|>|t^\prime|$, (b) semimetallic phase for $|t|=|t^\prime|$, and (c) metallic phase for $|t|<|t^\prime|$.
    Inset of (a) is the first BZ.
    (d) -- (f) Ribbon energy spectra of the $C_4$ model corresponding to the bulk phases in (a)-(c), showing robust edge states across all phases.
  }
\end{figure}

Figure 4(a) to (c) displays the bulk energy spectra for the unit cell of Fig.~3(b) for all symmetry-allowed parameters, i.e., $|t|>|t^\prime|$, $|t|=|t^\prime|$, and $|t|<|t^\prime|$.
For the bulk energy spectra shown in Figs.~4(a) to (c), we observe a double degeneracy at the $X=(\pi/a,0)$ point for all three cases.
This degeneracy can be regarded as the combined effect of time-reversal symmetry and $\mal$.
In the case of $|t|>|t^\prime|$, we regard it as the insulating phase due to the complete band gap between the first pair and the second pair of degenerate bands.
Conversely, for the case of $|t|<|t^\prime|$, the presence of a band overlap indicates a metallic phase, characterized by a lack of a band gap within the first four bands.
The intermediate case of $|t|=|t^\prime|$ is a semi-metallic phase characterized by a triple touching point at $\Gamma$.

Figures 4(d) to (f) display the ribbon energy spectra for the three cases of insulating, semimetallic, and metallic phases, respectively. As expected, all three cases exhibit distinct edge states due to nonvanishing electric polarization~\cite{supplement}.
These edge states emerge when the occupied number of bands is $2n$ for $ n=1,3$, and disappear when the number of occupied bands is $4n$ for $n=1,2$, due to the quantization of the total electric polarization at $e/2$.
The appearance of topological edge states for all three cases across insulating, semi-metallic and metallic phases confirms the BSI topological nature of the edge states induced by $\mathcal{L}$.
It is noted that these edge states are robust to the modification of terminations, such as removing a half unit cell for the edge sites as confirmed and seen next.
Intuitively, the $C_4$ model here can be regarded as two copies of the 2D SSH model in different topological phases connected by the $\mal$, which always yields nontrivial electric polarization.

\subsection{M{\"o}bius Edge States and Twisted Wilson Loops}
\begin{figure}[b]
  \leavevmode
  \begin{center}
    \leavevmode
    \includegraphics[clip=true,width=0.99\columnwidth]{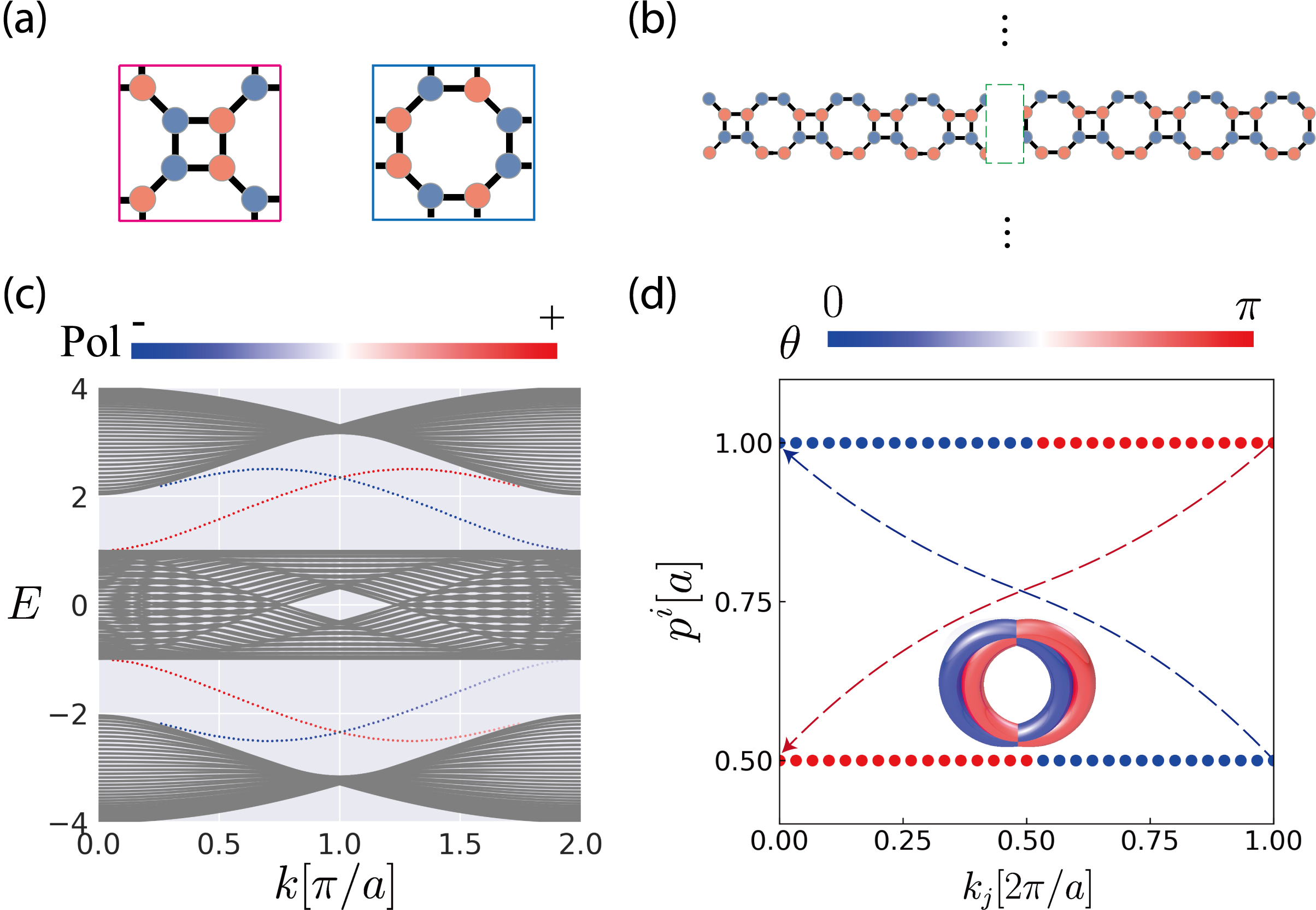}
  \end{center}
  \caption{
    (a) Two conjugated unit cells of the $C_4$ model connected by $\mbf{L}$ in ribbon structures, each possessing $\mal$ symmetry.
    (b) Conjugate junction formed by the two ribbons composed of the conjugated unit cells shown in (a).
    (c) Energy spectrum of the conjugate junction in (b), where gray represent the bulk states, blue and red represent edge states with opposite polarization. The ribbon is periodic along $y$-direction.
    (d) Wilson loop $p^i(k_j)$ of the $C_4$ model of first two bands for $|t|>|t^\prime|$ along the direction $i=x,y$. The colors indicate the angle of $\mal$ for two eigenstates of the Wilson loop, which swap at $k_j=\pi/a$.
    Inset is a Villarceau circle indicating the topological structure of the Wannier complex.
  }
\end{figure}

As discussed in the 1D case, a \Mob strip structure of wavefunctions emerges in momentum space due to the $\mal$ symmetry.
However, in the ribbon spectrum as displayed in Fig.~4(d) to (f) of the $C_4$ model, only one edge state per gap and termination is observed.
This is because of the specific $\mbf{L}=(a/2,a/2)$ adopted here.
Note that for a ribbon geometry of the $C_4$ model, $\mbf{L}$ does not map the ribbon onto itself; instead, it maps to a conjugated structure due to the open boundary, as displayed in Fig.~5(a).
The conjugate structure has exactly the same bulk and ribbon energy spectra as its partner, confirming the previouss discussed insensitivity of edge states to terminations.
By joining these two conjugated structures together as displayed in Fig.~5(b), \Mob edge states emerge as expected.

Figure 5(c) displays the energy spectrum of the conjugate junction with periodic boundary condition along $y$.
We observe a pair of edge states that are degenerate at $k=\pi/a$.
Essentially, these two edge states switch their components defined by $\mal$ at the degenerate point as indicated by the blue and red colors in Fig.~5(c), similar to a \Mob strip.
The two poles of the polarization in two edge states are defined as $|v'_2\rangle=|v_1\rangle+i|v_2\rangle$ and $|v'_1\rangle=|v_1\rangle-i|v_2\rangle$ with $|v_1\rangle$ and $|v_2\rangle$ being the degenerate states at $\pi/a$.
Then the polarization ratio is given as $\text{Pol}=(P_1-P_2)/(P_1+P_2)$ with $P_i=\langle v'_i|\psi\rangle$ as seen in Fig.~5(c).

Considering the \Mob edge states induced by $\mal$, we also expect a twist structure in the Wilson loop spectrum of the $C_4$ model.
From which we can define a nontrivial Chern number in a spin-polarized fashion.
For the $C_4$ model, the nonzero elements of the $\mal$ matrix are $e^{i(k_x+k_y)}$ for translated sites between $b$ and $d^\prime$, $e^{ik_y}$ for translated sites between $c^\prime$ and $a$, $1$ for translated sites between $d$ and $b^\prime$, and $e^{ik_x}$ for translated sites between $c$ and $a^\prime$.
The Wilson loop $p^i(k_j)$ is given by the eigenvalues of the overlap matrix that
\begin{equation}
  W^i_{k_j}=\prod_{k_i=0}^{2\pi}\langle u(k_i,k_j)|u(k_i+dk,k_j)\rangle,
\end{equation}
where $i=x,y$, $|u\rangle$ is the periodic part of Bloch functions of occupied bands, and $p^i(k)=\text{arg}(w^i_k)/2\pi$ with $w^i_k$s the eigenvalues associated with eigenvectors $|v^i_k\rangle$s.
Then, for $p^i$, we define the polarization angle $\theta$ in terms of $\mal$ as $\theta(k)=\text{arg}(\langle v^i_k|\mal(0,0)|v^i_k\rangle)$.

Figure 5(d) displays the Wilson loop for the lowest two bands with colors blue and red indicating $\theta$.
From Fig.~5(d), we observe that there are two lines of $p^i$ guaranteed by $\mal$, which have electric polarization $1.0$ (equivalent to 0) and $0.5$ as expected.
Essentially, as revealed by $\theta$, the polarization of two lines form a twist at $k_j=\pi/a$, and the two branches of the twist are $\theta=\pi$ and $\theta=0$ as indicated by blue and red in Fig.~5(d).
For the same color, the Wilson loop has a nontrivial winding, such as $C^\theta=p^{i,\theta}(2\pi)-p^{i,\theta}(0)=\pm1/2$ for $\theta=\pi$ and 0.
Similar to the quantum spin Hall effect, we obtain a nontrivial Chern number in a spin-polarized fashion that
\begin{equation}
  C_\text{(4)}=C^\pi-C^0=1,
\end{equation}
which indicates a nested double-rings structure similar to Villarceau circles as shown by inset of Fig.~5(d).

\subsection{Intrinsic Higher-Order Topological States}
\begin{figure}[t]
  \leavevmode
  \begin{center}
    \leavevmode
    \includegraphics[clip=true,width=0.99\columnwidth]{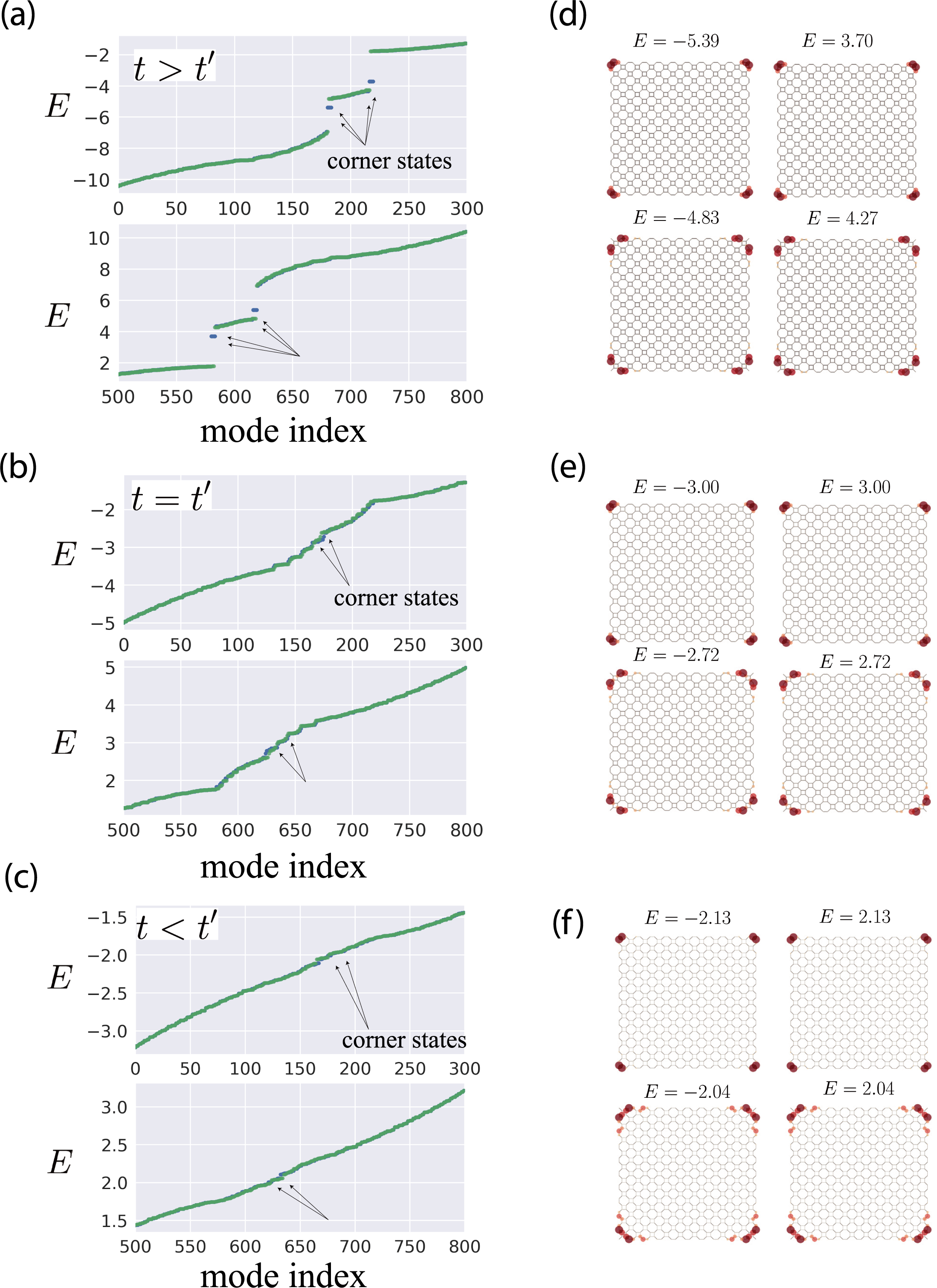}
  \end{center}
  \caption{
    Energy spectra for three cases: (a) $|t|>|t^\prime|$, (b) $|t|=|t^\prime|$, and (c) $|t|<|t^\prime|$ of the $C_4$ model with open boundary conditions along the $x$ and $y$ directions. The arrows indicate the corner states.
    (d) -- (f) Charge density distributions of the corner states for the three cases in (a)–(c).
  }
\end{figure}

Besides \Mob edge states, $\mal$ naturally induces higher-order topological states such as corner states~\cite{Lin2020}.
This is because of the following formula for the corner charge~\cite{Liu2019, Watanabe2020,Takahashi2021}
\begin{equation}
  Q_{xy}=\sum^m_ip^x_ip^y_i,
\end{equation}
where the summation takes over all components of the Wannier complex.
For the order-two case endowed with $\mbf{L}=(a/2,a/2)$, we have $Q_{12}=2xy+(x+y)/2+1/4$ with $\mbf{p}_1=(x,y)$ and $\mbf{p}_2=(x+1/2,y+1/2)$.
For $Q_{12}=0$, $x$ and $y$ have no rational solution.
Thus, corner states emerge intrinsically for order-two Wannier complex endowed with $\mbf{L}=(a/2,a/2)$ regardless of the value of $(x,y)$.

Figures 6(a) to (c) display the energy spectra of the $C_4$ model for finite samples with open boundary conditions along both the $x$ and $y$ directions.
Owing to the distinct geometric structures of the conjugated pair at the corners, the eigenenergies of the corner states in the conjugated pair are slightly different.
Figures 6(e) to (f) display the density profiles of the corner states for both structures of the conjugated pair in the three cases: $|t|>|t^\prime|$, $|t|=|t^\prime|$, and $|t|<|t^\prime|$.
It is noted that all the dangling sites in the finite sample have no effect on the formation of corner states.
For the gapless cases such as $t=t'$ and $t<t'$, these corner states are bound states in the continuum. 
To prevent hybridization with bulk states, additional symmetries are usually required.

\subsection{Glide Reflection in Energy-momentum Space without Fluxes}

\begin{figure}[t]
  \leavevmode
  \begin{center}
    \leavevmode
    \includegraphics[clip=true,width=0.99\columnwidth]{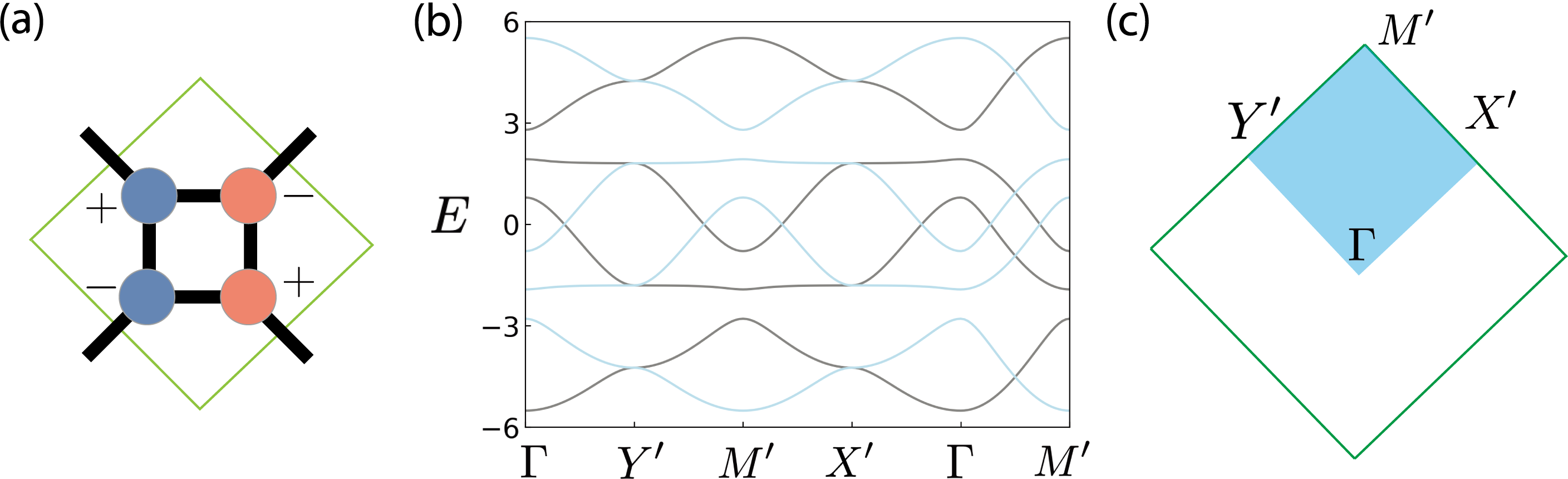}
  \end{center}
  \caption{(a) Primitive unit cell of the $C_4$ model with lattices $(a/2,a/2)$ and $(a/2,-a/2)$. ``$\pm$'' denotes two types of sublattices defined by $\mat{S}$.
    (b) Energy spectrum of $H_\mbf{k}$ (dark lines) and the shifted $E(k_x+2\pi/a,k_y)$ (light lines).
    (c) 1st Brillouin zone of the primitive cell in (a).
  }
\end{figure}
Here we show that $\mal$ induces a glide-reflection symmetry in energy--momentum space \emph{without} introducing a flux~\cite{Xiao2024}.
Considering $\mal$ switches sublattices in the unit cell, we can introduce a sublattice operator that $\mat{S}=\text{Diag}(1_A,-1_B)$, where $A$, $B$ are two types of sublattices that are connected by $\mbf{L}$.
Figure 7(a) displays the $A$- and $B$-type sublattices in the $C_4$ model, denoted by ``$\pm$''.
As $\mal$ switches $A$ and $B$, we have $\{\mal,\mat{S}\}=0$, with $\{\cdots\}$ indicating the anti-commutator.
For the primitive unit cell [Fig.~7(a) and the light diamond shape in Fig.~3(b)], $\mal=e^{ia(k_x+k_y)/2}$ and $\mat{S}\mal\mat{S}^{-1}=-e^{ia(k_x+k_y)/2}=e^{ia(k_x/2+k_y/2+\pi/a)}$.
This fact suggests that $\mat{S}$ shifts $\mbf{k}$ by $2\pi/a$ along the $k_x$ or $k_y$ direction.
Combining that $\mat{S}H_\mbf{k}\mat{S}^{-1}=-H_{\mbf{k}+\hat{x}2\pi/a}$, we obtain the following relation
\begin{equation}
  H_{\mbf{k}+\hat{x}2\pi/a}S|k_x,k_y\rangle=-ES|k_x,k_y\rangle,
\end{equation}s
where $H_k|k_x,k_y\rangle=E|k_x,k_y\rangle$.
Equation (12) yields a glide-reflection symmetry in energy--momentum space such that
\begin{equation}
  E(k_x+2\pi/a,k_y)=-E(k_x,k_y).
\end{equation}
Figure 7(b) displays the energy spectrum of $H_\mbf{k}$, where the light-blue and black lines are $E(k_x+2\pi/a,k_y)$ and $E(k_x,k_y)$, respectively. 
From Fig.~7(b) we observe that $E(k_x+2\pi/a,k_y)$ coincides with $-E(k_x,k_y)$ well.
Figure 7(c) displays the corresponding 1st Brillouin zone of the primitive unit cell of the $C_4$ model.

\section{Higher-order BSI topology wiht $C_2$ point-group symmetry}

In the $C_4$ symmetric case, we show that $\mbf{L}=(a/2,a/2)$ together with $C_4$ point-group symmetry fixes the total electric polarization of the Wannier complex at $(1/2,1/2)$, which yields a BSI topological system associated with \Mob edge states and intrinsic higher-order corner states. In the $C_2$ symmetric case, the degeneracy of the general Wyckoff position is two; thus, a first-order BSI topological system cannot exist.

In this section, we show that by adding asymmetric onsite potentials in the $C_4$ model, the BSI topological system devolves to a conventional topological system in the first-order, 
whose topological phase transitions are characterized by band inversion, accompanied with persisting higher-order topological states.

Let us first look at possible configurations of Wannier complex under $C_2$ symmetry constraint.
The four special Wyckoff positions of $C_2$ symmetric lattice are $1a$, $1b$, $1c$ and $1d$, as displayed in Fig.~8(a).
Similar to the $C_4$ case, there are three types of $\mbf{L}$: $(a/2,0)$, $(0,a/2)$, and $(a/2,a/2)$.
For $\mbf{L}=(a/2,a/2)$, the order of Wannier complex is two with different total electric polarization of configurations.
For zero total electric polarization, we have the Wannier complex $\mbf{p}_1=(-a/4,-a/4)$ and $\mbf{p}_2=(a/4,a/4)$, and for nontrivial total electric polarization, we have $\mbf{p}_1=(0,0)$ and $\mbf{p}_2=(a/2,a/2)$.
For which configuration of Wannier complex is taken in a specific model, it is determined by the band structure similar to the case of conventional topological systems.

\begin{figure}[t]
  \leavevmode
  \begin{center}
    \leavevmode
    \includegraphics[clip=true,width=0.99\columnwidth]{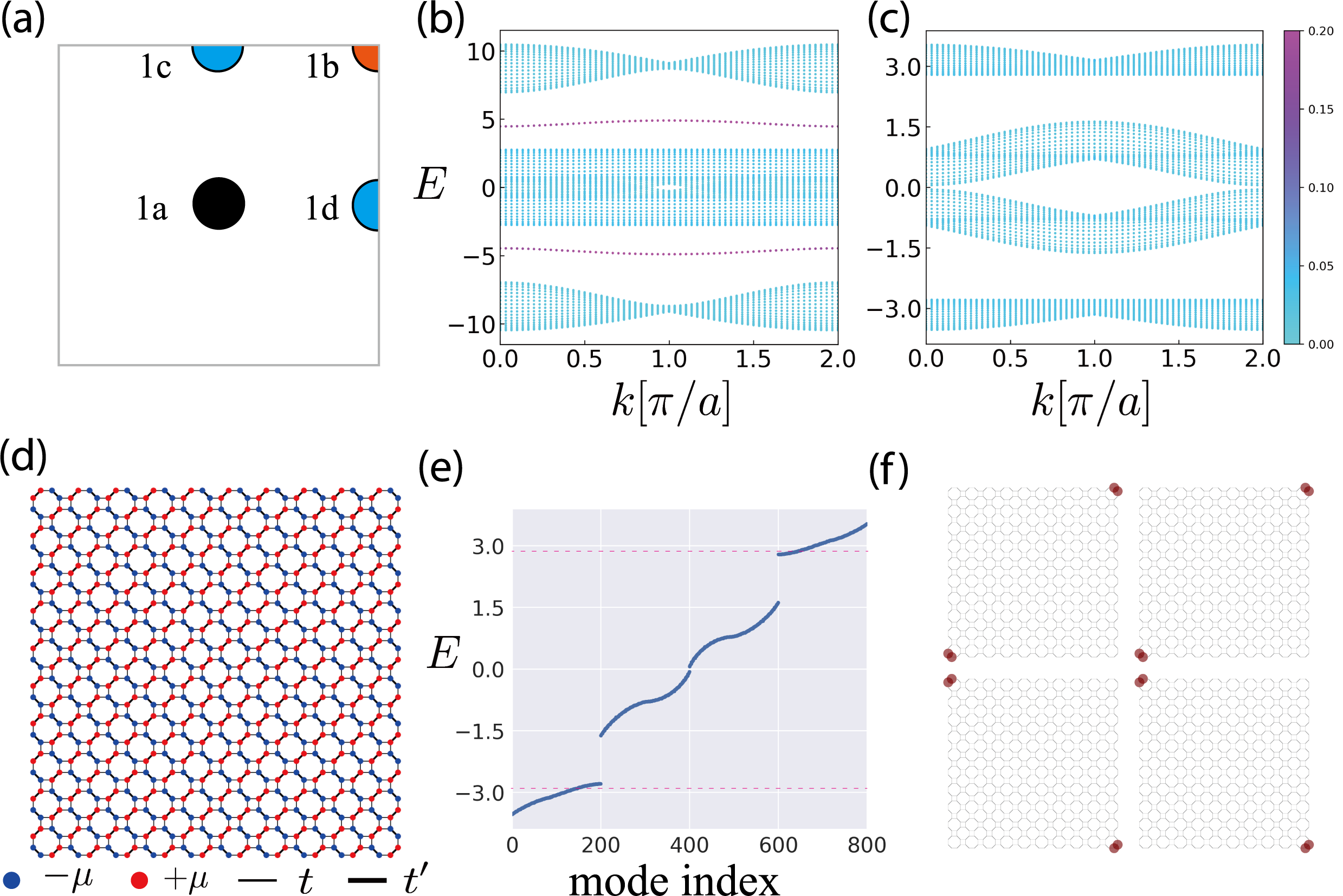}
  \end{center}
  \caption{(a) Special Wyckoff positions of $C_2$ lattice.
    (b) Ribbon spectrum of the $C_2$ model for $|t|>|t^\prime|$ and $\mu=1$.
    The color-map indicates the inverse participation ratio of wavefunctions at each $k$, where the darker color presents the edge states, and the lighter one are bulk states.
    (c) Similar to (b) for $|t|<|t^\prime|$ and $\mu=1$.
    (d) Schematic figure of a finite sample of $C_2$ model. The blue and red circles are for opposite onsite energies, and thick and thins lines are for $t$ and $t^\prime$ hoppings.
    (e) Energy spectrum of the finite sample in (d) for $|t|<|t^\prime|$ and $\mu=1$.
    The dashed lines indicate those eigenenergies of corner states.
    (f) Charge density distributions of corner states marked in (e).
  }
\end{figure}

To confirm this point, we take the previous $C_4$ model with a nonzero onsite potential $\mu$ as an example, which reduces the $C_4$ model to a $C_2$-symmetric one.
Figures 8(b) and (c) display the energy spectrum of ribbon structures for $|t|>|t^\prime|$ and $|t|<|t^\prime|$.
From Figs.~8(b) and (c) we see that edge states only emerge for the case $|t|>|t^\prime|$ and are absent for $|t|\le |t^\prime|$,  indicating a first-order topological phase transition driven by tunning hopping amplitude.

\subsection{Pure Quadrupole Phase}
As discussed already, for order-two Wannier complex endowed with $\mbf{L}=(a/2,a/2)$, corner states appear intrinsically due to finite corner charge.
This argument also applies to the $C_2$ model, even in the absence of edge states for $|t|\le |t^\prime|$.
Figure 8(d) displays the schematic figure of the $C_2$ model with open boundary condition along both $x$- and $y$-directions.
It is noted that for the conjugated structure, the following shown result is similar.
Figure 8(e) displays the energy spectrum of the finite sample in Fig.~8(d) for $|t|\le |t^\prime|$, and the corresponding corner states are marked by dashed lines.
From Fig.~8(e) we see that there are no edge states within the band gaps, but there are corner states.
Figure 8(f) displays the charge density distributions of the four corner states, which have opposite energies.
Thus, in the $C_2$ model for $|t|\le |t^\prime|$, it is a pure quadrupole phase, characterized by corner states in the absence of edge states.

\section{Cases with $C_3$ and $C_6$ point-group symmetries}

In the $C_4$ and $C_2$ models, we focus on the order-two Wannier complex.
Here we extend our discussion to other orders of Wannier complexes, such as three and six orders in $C_3$ and $C_6$ symmetric lattices.
In general, we have $\mal^n=e^{i\mbf{k}\cdot\mbf{a}}$.
For $n=3$, $\mal$ can be written as
\begin{equation}
  \mal_{(3)}=
  \begin{pmatrix}
    0&e^{i\mbf{k}\cdot\mbf{a}} \otimes 1_M &0 \\
    0&0&1_M\\
    1_M&0&0
  \end{pmatrix},
\end{equation}
where the unit cell is the tripled one with total $3M$ sublattices of three types, and the bases are those $|A_{1\dots M}\rangle$,$|B_{1\dots M}\rangle$,$|C_{1\dots M}\rangle$ connected by $\mal_{(3)}$, and each type have $M$ sublattices.

\subsection{$C_3$ model}

\begin{figure}[t]
  \leavevmode
  \begin{center}
    \leavevmode
    \includegraphics[clip=true,width=0.99\columnwidth]{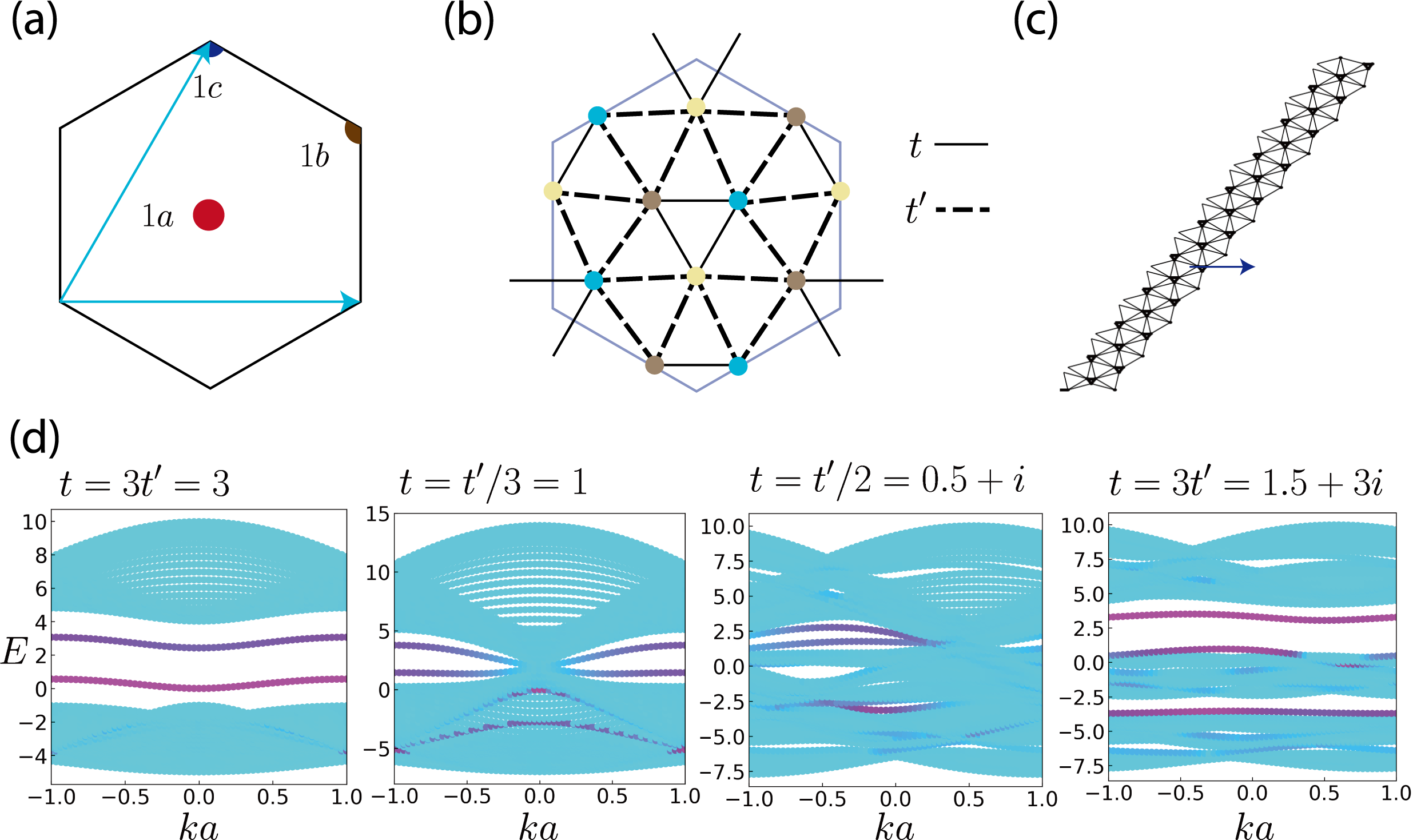}
  \end{center}
  \caption{(a) Special Wyckoff positions of $C_3$ symmetric lattice.
    The arrows are lattice vectors.
    (b) Schematic of unit cell of the $C_3$ lattice model with $\mbf{L}=(a/2,a/2\sqrt{3})$.
    Solid and dashed lines represent two types of hoppings.
    Colors are an eye guide for equivalent sites connected by $\mbf{L}$.
    (c) Ribbon structure of the $C_3$ model with periodic boundary condition along $(1,0)$ as indicated by the arrow.
    (d) Energy spectra of the ribbon structure in (c) for different parameters.
    The color-map shows the inverse participation ratio of the wavefunctions at each $k$ point, where light indicates bulk states and dark indicates edge states.
  }
\end{figure}

In the case of $\mal^3=e^{i\mbf{k}\cdot \mbf{a}}$, the order of Wannier complex is multiple of three.
For a concrete example, we consider a $C_3$ symmetric lattice, whose special Wyckoff positions are displayed in Fig.~9(a).
There are three special Wyckoff positions $1a$, $1b$ and $1c$ as marked in Fig.~9(a) with arrows indicating the lattice vectors.
We consider a $\mbf{L}$ connecting $1a$ and $1b$ (or $1c$), and construct a $C_3$ model, whose schematic of unit cell is displayed in Fig.~9(b).
There are three types of sublattices connected by $\mbf{L}$, and two different kinds of hoppings, which can be regarded as first-nearest hopping $t$ and second-nearest hopping $t^\prime$.

Considering the $C_3$ point-group symmetry, the order of Wannier complex equipped with $\mbf{L}=(a/2,a/2\sqrt{3})$ is three.
The topological characterization of order-three Wannier complex is given by $\text{Sym}^3(T^2)$, where $T^2$ is a 2D torus.
$\text{Sym}^3(T^2)$ has a $\mathbb{CP}^2$ line bundle pinning to point $p$ on $T^2$, which is topologically nontrivial.
This suggests that the $C_3$ model is another example of a BSI topological system.

To confirm the nontrivial topology of the $C_3$ model, we construct a ribbon structure that is periodic along $(1,0)$ as displayed in Fig.~9(c).
For different hopping parameters including those breaking time-reversal symmetry, we all observe the emergence of edge states displayed in Fig.~9(d).
Especially, for $|t|<|t^\prime|$ it is a topological semi-metallic phase characterized by Dirac-cone-like edge states.
It is noted that for ribbons periodic along other directions, like the armchair direction $(0,\sqrt{3})$, the ribbon spectra are similarly possessing edge states.
The emergence of edge states in various ribbon structures for different parameters demonstrates the nontrivial topological nature of the $C_3$ model.
\begin{figure}[t]
  \leavevmode
  \begin{center}
    \leavevmode
    \includegraphics[clip=true,width=0.99\columnwidth]{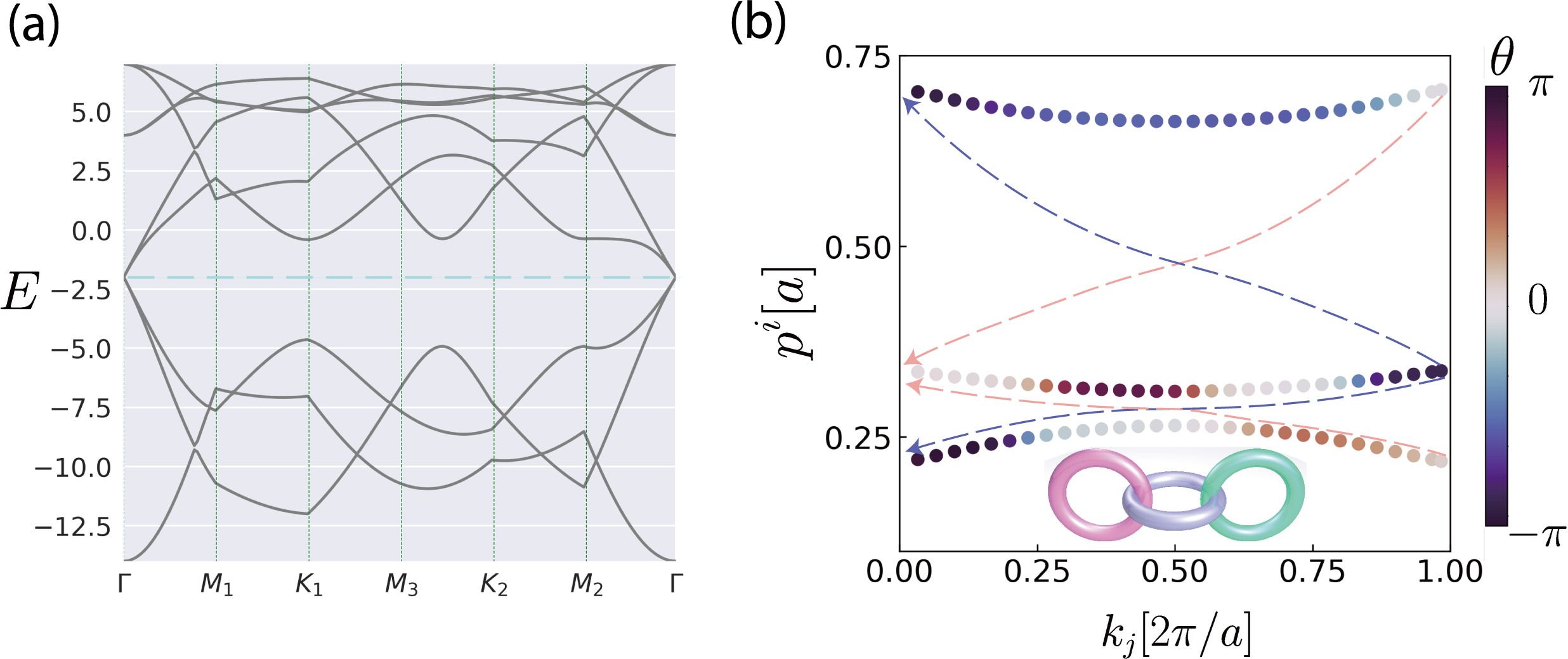}
  \end{center}
  \caption{(a) Bulk energy spectrum of the $C_3$ model for $t=-1$ and $t^\prime=-3$.
    The horizontal dashed line indicates the position of fermi energy.
    (b) Wilson loops of the first three energy bands of the $C_3$ model shown in (a).
    The color-map indicates the angle of $\mal_{(3)}$ operator for the Wilson loop eigenstates.
    The dashed arrow lines show the connection between Wilson loops by the continuous color-map.
    Inset is three linked circles indicating the topological structure of the Wannier complex.
  }
\end{figure}

Distinct from the order-two case, order-three Wannier complex features richer topological structures, for example, twists among three Wilson loops.
We consider the parameter choice $t=-1$ and $t^\prime=-3$ in the $C_3$ model and set the Fermi energy for the first three bands, where the valence and conduction bands touch at $\Gamma$.
Figure 10(a) displays the bulk energy band structure for the chosen parameters, where the dashed line indicates the position of the Fermi energy.
Because of the three types of sublattices, there is no chiral symmetry.
Figure 10(b) shows the Wilson loops for the three occupied states, with the color-map indicating the angle $\theta$ of the $\mal_{(3)}$ operator for each Wilson loop eigenstates, similar to Fig.~5(d).
From Fig.~10(b) we observe three Wilson loops as finite values and they are twisted together in $\theta$-polarized fashion as indicated by the dashed arrow lines.
This twisted structure of Wilson loops can be characterized by the following Chern number in the order of connections indicated by $\theta$ that
\begin{equation}
  C_{(3)}=2\times[p_3(2\pi/a)-p_2(0)+p_2(2\pi)-p_1(0)]=1,
\end{equation}
which indicates its nontrivial topology, as shown by the three nested rings in the inset of Fig.~10(b).

\subsection{$C_6$ model}
The $C_6$-symmetric lattice is similar to the $C_3$ one and has three special Wyckoff positions.
Under the constraint of $C_6$ point-group symmetry, the order of Wannier complex could be three and six for the same $\mbf{L}=(a/2,a/2\sqrt{3})$ of the $C_3$ symmetric case.
This is different from other point groups, which have same order of Wannier complex for a fixed $\mbf{L}$.
Thus, a $C_6$-symmetric model equipped with $\mal$ cannot be a BSI topological system.
For brevity, we simply skip the discussion of the $C_6$ model.

\section{Experimental Proposals}
With the study of $\mal$ symmetry in different point groups, we now turn to possible experimental verification of our toy models in realistic systems.
Here we consider two types of crystalline systems: one is a dielectric photonic crystal, and the other is a 2D solid-state material.
In both cases, we focus on the unique topological properties brought by $\mal$ together with point-group symmetry, such as BSI $\mbf{P}$, $\theta$-polarized edge states, and corner states.

\subsection{BSI Topology in a Dielectric Photonic Crystal}
\begin{figure}[b]
  \leavevmode
  \begin{center}
    \leavevmode
    \includegraphics[clip=true,width=0.99\columnwidth]{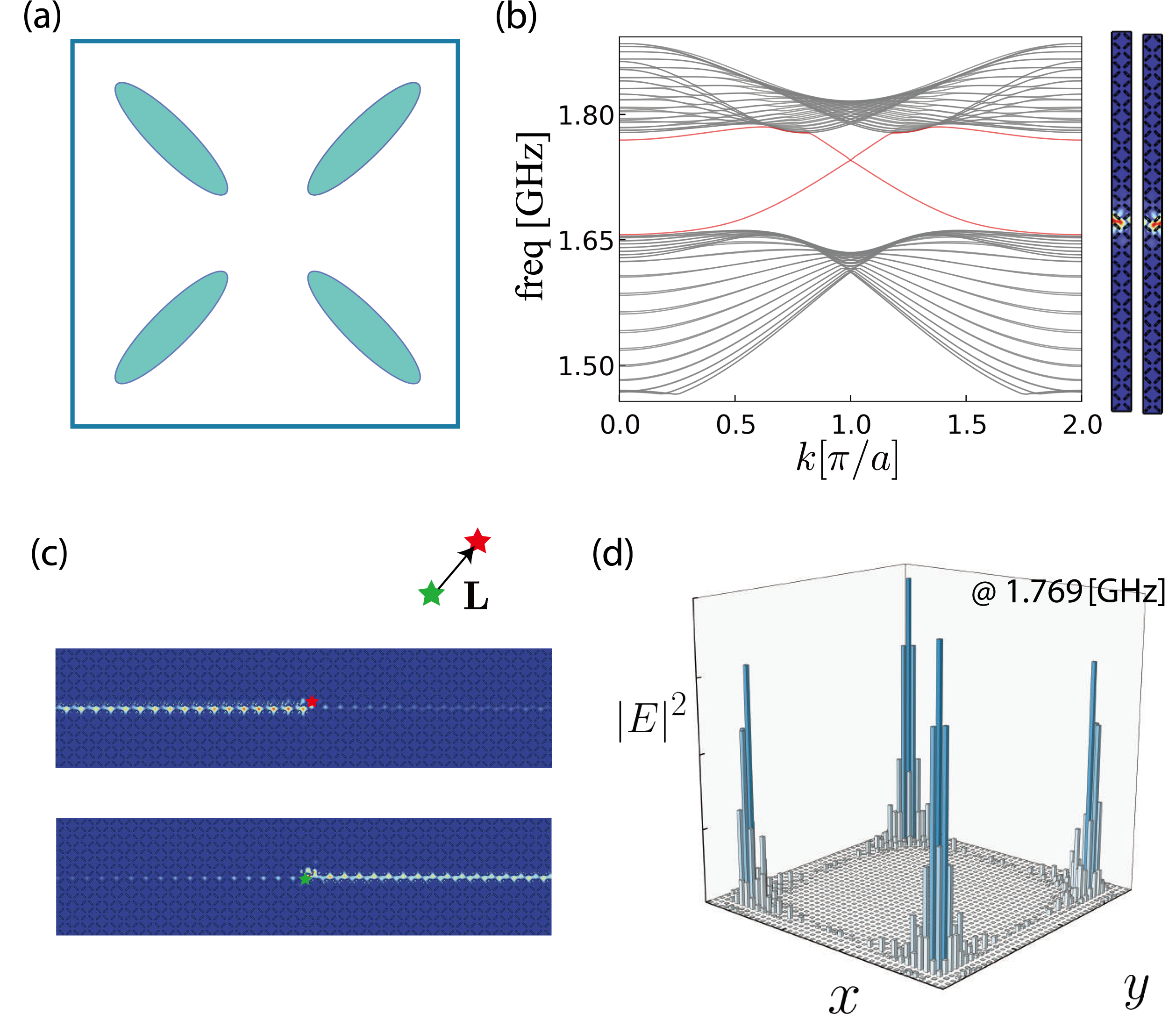}
  \end{center}
  \caption{ (a) Unit cell of the dielectric photonic crystal with $\mathcal{L}$ symmetry. Dark ellipses represent dielectric materials of $\varepsilon=11$.
    (b) Ribbon spectrum of the topological interface structure of the photonic crystal.  Inset are the degenerate topological interfacial states for $k=\pi/a$.
    (c) Electric field profile of the corner state at eigenfrequency 1.769[GHz] of the dielectric photonic crystal.
    (d) Selective excitation of edge states with opposite propagation in a topological interface structure of the dielectric photonic crystal.
  The sources for different propagation directions are spaced by $\mbf{L}$.}
\end{figure}

We construct an artificial crystalline system using dielectric cylinders to realize the $C_4$ model discussed in Sec.~III, and study its relevant properties through full-wave finite-element simulations using COMSOL.
Owing to the absence of quantum fluctuations and strong correlations in classical electromagnetic systems, the results of finite-element simulations should coincide with experimental results as long as the sample fabrication quality is sufficiently good.

Figure 11(a) displays the schematic of the design of the dielectric photonic crystal mimicking the $C_4$ model, where the dark ellipses represent a material of dielectric constant $\varepsilon$, i.e., $\varepsilon=11$, the background is simply air, the centers of four ellipses are $(d\sin(\pm\pi/4),d\cos(\pm\pi/4))$ with $d=0.35a$, and the major and minor axes of ellipses are $0.4a$ and $0.1a$, respectively.
From Fig.~11(a), we observe the designed photonic crystal holds $\mbf{L}=(a/2,a/2)$ and $C_4$ point-group symmetry.
Similar to the $C_4$ model, there are two conjugated unit cells of the designed photonic crystal related by $\mbf{L}$.
Figure 11(b) displays the energy spectrum of the ribbon constructed by two types of unit cells featured by the conjugated junction.
We observe a pair of edge states that is degenerate at $k=\pi/a$ within the photonic band gap in Fig.~11(b).
These degenerate edge states emerge universally for occupied number $n=2,6,10,\dots$ irrespective of specific photonic band structures~\cite{supplement}.
The inset of Fig.~11(b) displays the profile of the out-of-plane electric field, which is concentrated near the conjugate junction.

Owing to the connection between the two edge states by $\mbf{L}$ shown in Fig.~11(b), we expect a switching behavior of edge states by shifting the excitation source of $\mbf{L}$ without changing the frequency or light polarization.
Figure 11(c) shows the harmonic excitation behavior of the finite photonic crystal sample, where the conjugated junction is built along the $x$-direction.
From Fig.~11(c) we observe that, only by shifting the point source (indicated by the green and red stars) by $\mbf{L}$, the propagation direction of the excited edge state reverses, confirming our previous discussion.
Figure 11(d) displays the electric field profile of the topological corner state for the photonic crystal, where the electric field is well-localized around the four corners symmetrically, and the eigenfrequency is within the band gap of bulk states.

\subsection{Electric BSI topological platform: Octa-graphene}

\begin{figure}[t]
  \centering{}
  \includegraphics[clip=true,width=0.99\columnwidth]{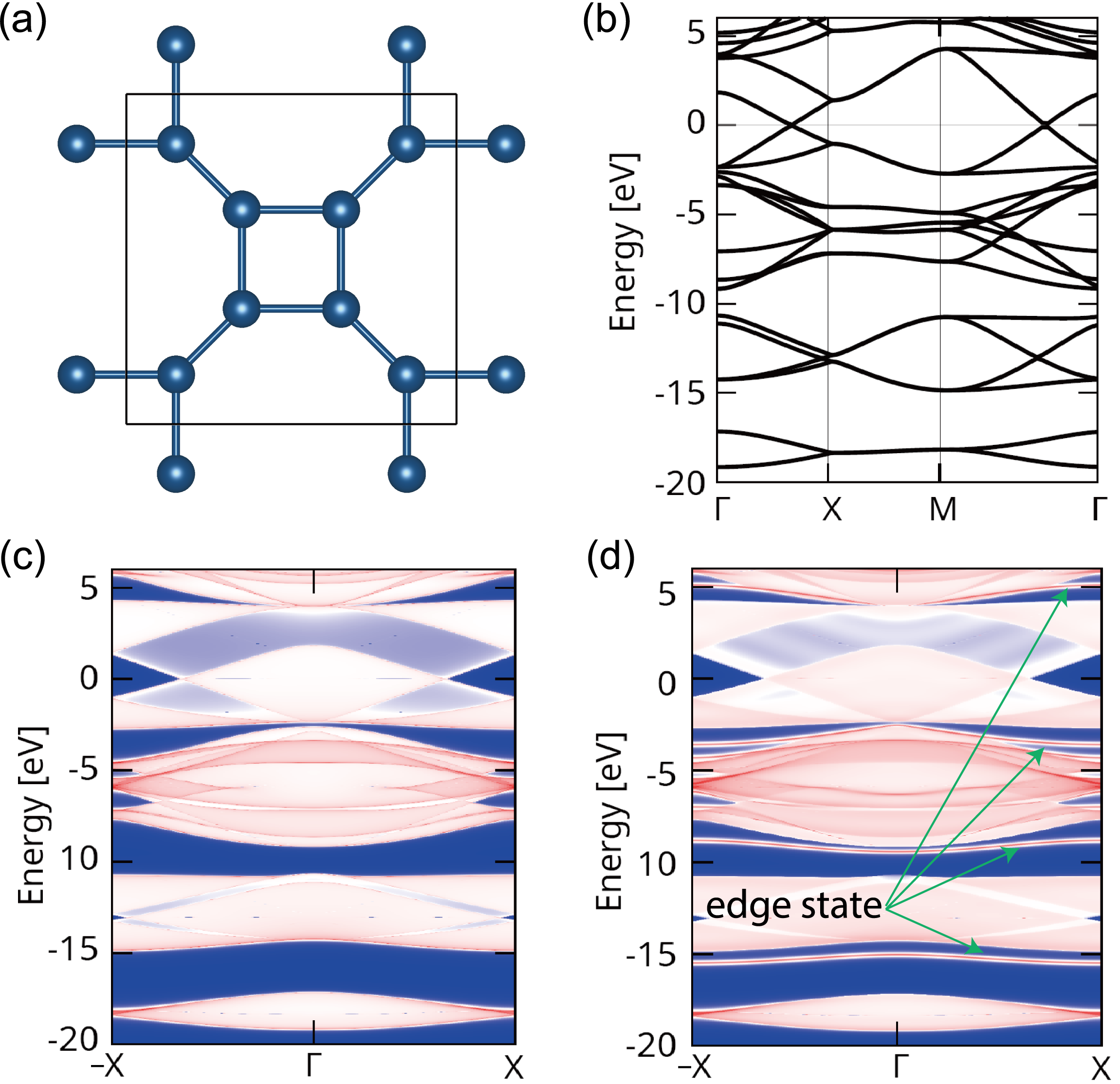}
  \caption{Topological electronic structure of octa-graphene. (a) Two-dimensional crystal structure of octa-graphene, composed of square carbon rings interconnected by octagonal rings. A $\sqrt{2}\times\sqrt{2}$ enlarged unit cell is shown. (b) Band structure of octa-graphene. (c) and (d) show the bulk and surface band structures of a ribbon, respectively. Edge states in the surface band structure are highlighted by arrows.
  \label{fig1:octagr} }
\end{figure}

A metastable form of graphene has been predicted, featuring a periodic planar sheet of sp$^2$-bonded carbon atoms arranged into linked square rings interconnected by octagonal carbon rings~\cite{sheng2012} (Fig.~\ref{fig1:octagr}(a)).
This \emph{octa$-$graphene} structure is energetically more favorable than graphyne and graphdiyne.
The lattice of octa$-$graphene shares the same underlying crystal symmetry as the toy tight$-$binding model illustrated in Fig. 3(b), but it incorporates multiple atomic orbitals per site, enabling a richer electronic structure.

The band structure of octa$-$graphene was computed using first$-$principles calculations performed with the Vienna Ab initio Simulation Package (VASP)~\cite{vasp}, and the resulting energy bands are displayed in Fig.~\ref{fig1:octagr}(b). The monolayer exhibits a metallic band structure with a nodal-line crossing at the Fermi energy that intersects the high-symmetry paths $\Gamma$--$M$ and $\Gamma$--$X$. These bands organize into distinct groups, with the number of bands in each group following the sequence 2--4--8--4 from low to high energy. Based on this grouping, one anticipates the emergence of topologically protected edge states within the band gaps separating adjacent groups---specifically, in those gaps where the cumulative number of bands below the gap is not a multiple of four. This expectation is confirmed by an explicit calculation of the surface states for a finite ribbon of the material.

We computed both the bulk and edge band structures for a ribbon geometry. The bulk bands were obtained under fully periodic boundary conditions, whereas the edge (surface) bands were calculated with periodic boundary conditions imposed along the ribbon axis and open boundary conditions applied across its width, enabling a direct comparison between bulk and boundary electronic states. The corresponding results are shown in Fig.~\ref{fig1:octagr}(c) (bulk) and Fig.~\ref{fig1:octagr}(d) (edge), respectively. Clear edge-localized bands appear within the predicted energy gaps in Fig.~\ref{fig1:octagr}(d), providing direct evidence of nontrivial topology. These findings validate the symmetry-based arguments presented in this work and demonstrate that the proposed topological criterion remains robust even for a realistic, multi-orbital electronic structure of a condensed-matter system.

\section{BSI topology in a 3D case}
\begin{figure}[t]
  \leavevmode
  \begin{center}
    \leavevmode
    \includegraphics[clip=true,width=0.99\columnwidth]{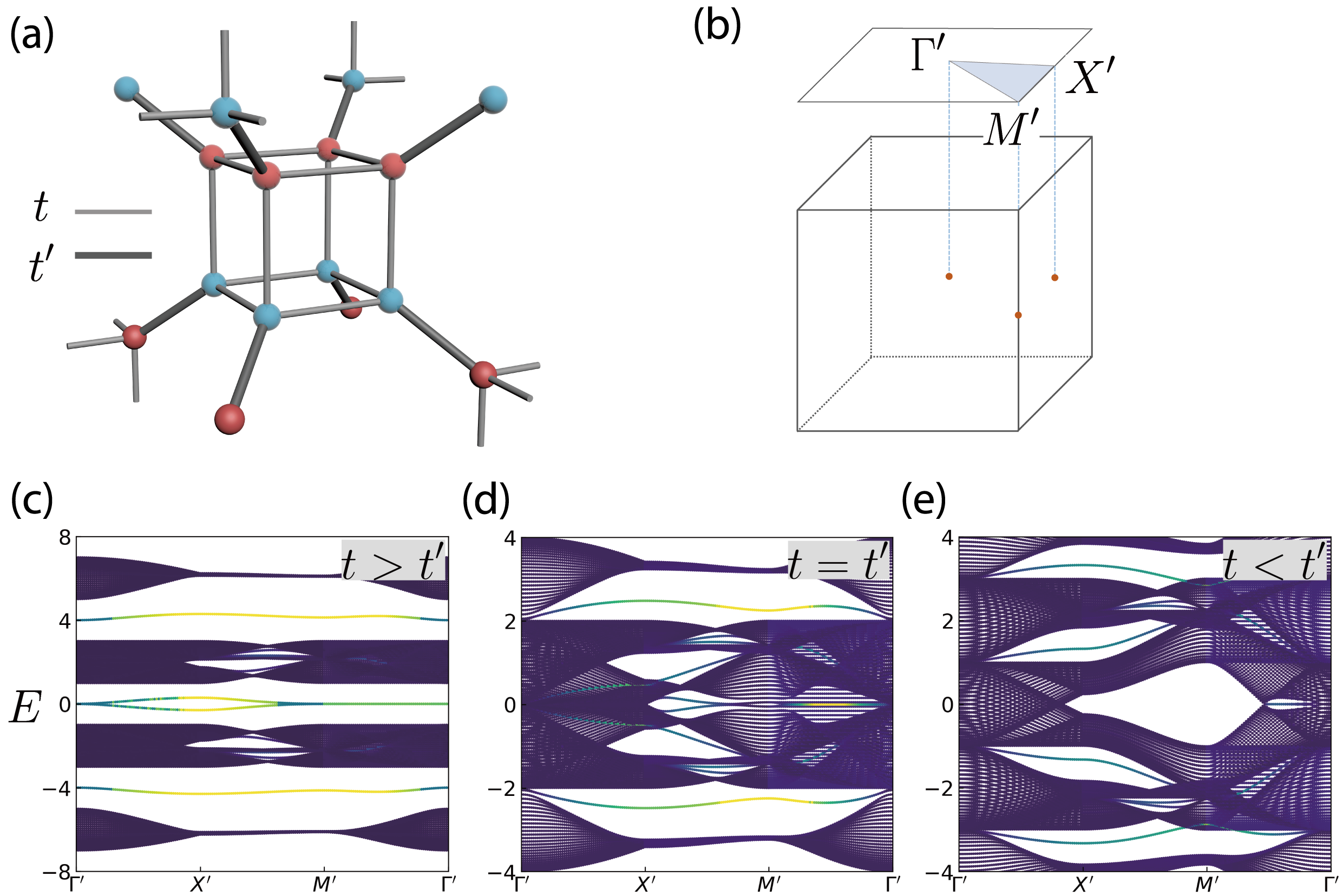}
  \end{center}
  \caption{ (a) Schematic of the unit cell of the BSI 3D model with $\mbf{L}=(a/2,a/2,a/2)$. Dark and light sticks represent two types of hoppings, $t$ and $t^\prime$. Colors of sublattices serve as an eye guide for equivalent sublattices connected by $\mbf{L}$.
    (b) 1st BZ of the BSI 3D model and its corresponding projected BZ in 2D.
    (c) to (e) Energy spectra of slab structures that are periodic along the $x$ and $y$ directions for three different cases: $|t|>|t^\prime|$, $|t|=|t^\prime|$, and $|t|<|t^\prime|$.
  The color maps show the inverse participation ratio of the wavefunctions at each $k$ point, where dark indicates bulk states and light indicates edge states.}
\end{figure}
Finally, we present an example of a 3D system with BSI topological properties.
In general, BSI topological materials should be sought in nonsymmorphic space groups, such as the space group $Pmmm$ where the high-$T_c$ superconductor yttrium barium copper oxide resides.
Here we construct a BSI topological system by extending the $C_4$ model along the $z$ direction.
Figure 13(a) displays a schematic of the unit cell of the 3D BSI model, where there are 16 sublattices in the unit cell, and it is equipped with $\mbf{L}=(a/2,a/2,a/2)$ in a cubic lattice structure.
Similar to the $C_4$ model, the order of Wannier complex is two, but here the base space is $T^3$ rather than $T^2$.
Figure 13(b) displays its corresponding first BZ.
Figures 13(c) to (e) display the slab spectra of the 3D BSI model for different hopping parameters, where surface states emerge universally.
As expected, in all parameters shown, there exist topological edge states, confirming its BSI topological nature.
It is noted that, due to the degrees of freedom of atomic orbitals, the surface states show richer band structures than the 2D cases.

\section{Summary}
In summary, we present a new framework for topological phases that is independent of a material's microscopic band structure.
This result opens the door to discovering novel topological materials in which symmetry, rather than a specific band structure, dictates the presence of robust topological edge and corner states.
The band-structure-independent topology is achieved through a band degeneracy enforced by the fractional translational symmetry $\mathcal{L}$, which necessitates a description in terms of a multiband Wannier complex, marking a shift in viewpoint from individual bands to symmetry-enforced multiband objects for determining topological properties.

The key signature of this topology is the universal quantization of electric multipoles, which in turn guarantees the presence of directional topological edge states and localized corner states. Crucially, since this mechanism is insensitive to hopping parameters and persists across insulating and metallic phases, it establishes a new paradigm for discovering topological materials suitable for robust waveguiding. We further demonstrate the experimental feasibility of this concept through a dielectric photonic crystal and a solid-state electronic material that exhibit all the hallmark properties predicted by our theory.

F. Liu thanks S. Iwamoto, C. K. Zhang, and R. Malte for useful discussions.

\bibliography{references}
\end{document}